\documentclass[apjl,iop]{emulateapj}
\pdfoutput=1
\usepackage{aas_macros}

\usepackage{float}
\usepackage{ctable}

\begin{document}

\title{Effects of the neutrino mass splitting on the non-linear matter power spectrum}
\author{Christian Wagner\altaffilmark{1}, Licia Verde\altaffilmark{2,1},  Raul Jimenez\altaffilmark{2,1}}
\altaffiltext{1}{Institut de Ciencies del Cosmos (ICC), Universitat de Barcelona (IEEC-UB), Marti i Franques 1, E08028 Barcelona, Spain}
\altaffiltext{2}{ICREA, Instituci\'o Catalana de Recerca i Estudis Avan\c{c}ats.}

\begin{abstract}
We have performed cosmological N-body simulations which include the effect of the masses of the individual neutrino species. The simulations were aimed at studying the effect of different neutrino hierarchies on the matter power spectrum. Compared to the linear theory predictions, we find that non-linearities enhance the effect of hierarchy on the matter power spectrum at mildly non-linear scales. The difference between the different hierarchies is about 0.5\% for a sum of neutrino masses of  $0.1$eV. Albeit this is a small effect, it is potentially measurable from upcoming surveys.
In combination with neutrinoless double-$\beta$ decay experiments, this opens up the possibility of using the sky to determine if neutrinos are Majorana or Dirac fermions.  
\end{abstract}


\section{Introduction}
Neutrino oscillation experiments have demonstrated that neutrinos have non-zero mass. Measurements of the flavor changing oscillations have provided a difference in the squares of the masses of the lightest and heaviest mass eigenstates $\Delta m^2 \simeq (0.05{\rm eV})^2$, yielding therefore a lower limit on the total neutrino mass ($\Sigma=\sum m_{\nu_i}$).  On-going and forthcoming ground based neutrino experiments  are sensitive to neutrino flavor and to the nature of neutrino mass (Dirac or Majorana) but are only sensitive to the absolute mass scale for large masses.
On the other hand, cosmological probes are blind to flavor but sensitive to the absolute  neutrino mass scale and 
 there has been recently  significant progress in constraining the sum of neutrino masses from cosmological observations.
 Massive neutrinos affect the observed matter power spectrum: their free-streaming damps the small-scale power and modifies the shape of the matter power spectrum below the free-streaming length \citep{Doro,HuEisensteinTegmark,TakadaKomatsu,0709.0253}. Up-to-date only upper limits have been obtained. However, the constraints are getting tighter --the current  limits on the total mass are $ \lesssim 0.3$eV (e.g., \cite{Lahavmassnu,Reid,Komatsu,Saito,Riemer,Roland})-- and  closer to the experimental limits derived from accelerator, reactor, solar, and atmospheric neutrino oscillations (see the reviews by \cite{pastor,concha} and references therein). Detecting the effect of neutrino masses on cosmological structure and measuring the neutrino mass scale is well within the reach of upcoming cosmological surveys (e.g., \cite{TakadaKomatsu,HannestadWong,Kitching_nu,LSST,Hannestadreview, LahavDES,Reid,Jimenez,Carbone} and references therein).
 
\begin{figure}[htb]
\begin{center}
\includegraphics[angle=0,width=0.45\textwidth]{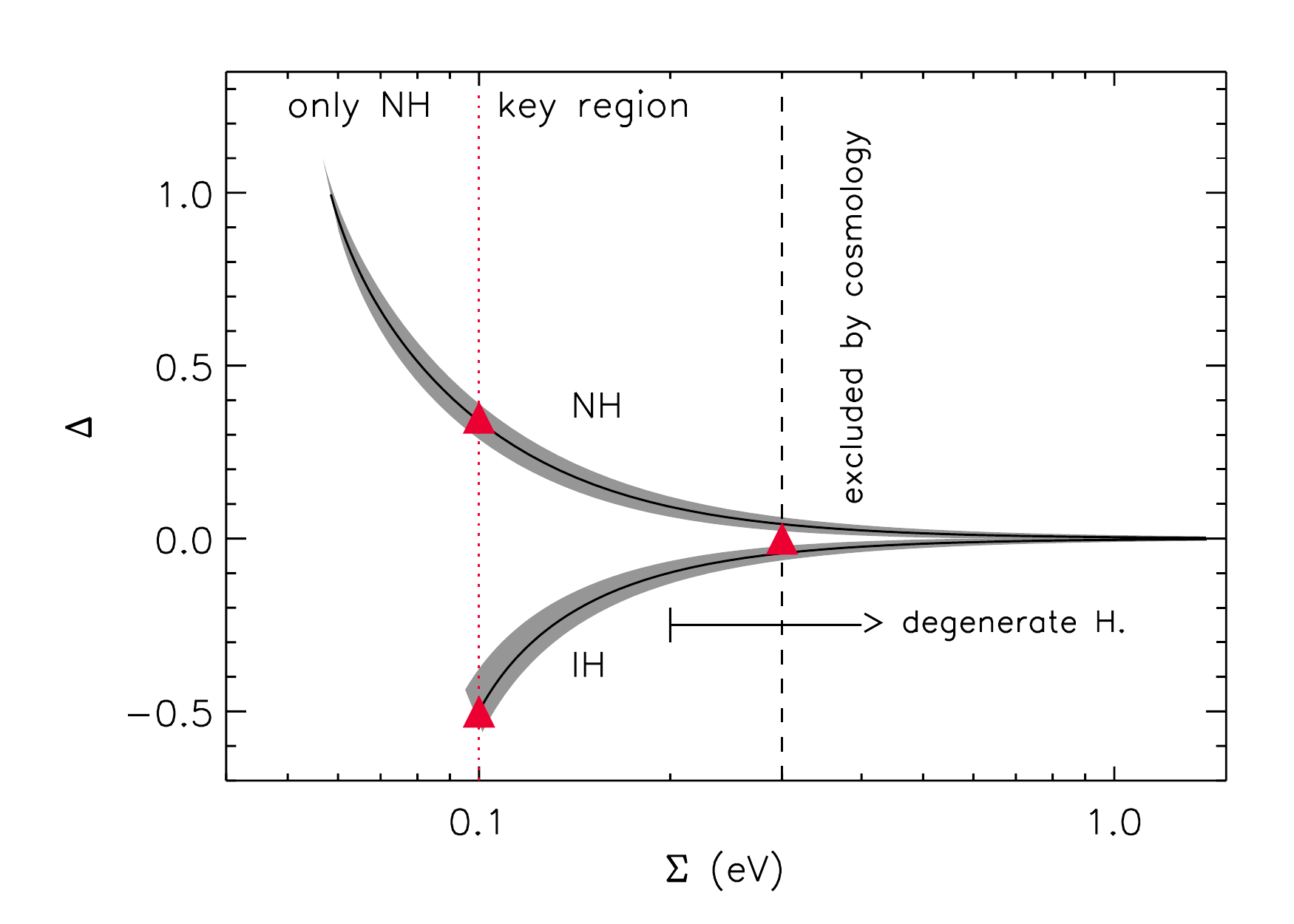}
\end{center}
\caption
{Constraints on the mass splitting from neutrino oscillations (shaded regions) and total neutrino mass from cosmology (vertical  dashed line) in the parameter space defined by the sum of neutrino masses $\Sigma$ and the mass splitting parameter $\Delta$ characterizing the hierarchy.  The key region where it is interesting to determine $\Delta$ is  $0.1 \rm{\, eV} <\Sigma<0.3 \rm{\, eV}$. The triangles indicate the cases studied in this paper for normal and inverted hierarchies (at $\Sigma=0.1$eV) and for the degenerate hierarchy (at $\Sigma=0.3$eV). Plot adapted from Fig.~1 of \citet{Carbone}.}
\label{fig:delta_sigma}
\end{figure}

The neutrino mass splitting required to explain the oscillation results  implies  that for three neutrino species there are two possible hierarchies in the neutrino mass spectrum: normal hierarchy (NH) with two light states and one heavy one  and a total mass $\Sigma \gtrsim 0.05$eV; inverted hierarchy (IH) with two heavy states and one light one with $\Sigma \gtrsim 0.1$eV. If the absolute mass scale is much higher than the mass splitting then the mass hierarchy does not matter and this case is referred to as degenerate mass spectrum. The degenerate hierarchy  however is under pressure from observations \citep{Lahavmassnu,Reid,Concha2}; see Fig.~\ref{fig:delta_sigma}, where we have introduced the  neutrino mass splitting parameter $\Delta$ relating  the  lightest ($m$) and heaviest ($M$) neutrino masses following \citet{Jimenez}  \footnote{Since one mass splitting is much larger than the other, for cosmological applications we can safely ignore the small mass splitting.}
\begin{eqnarray}
{\rm NH:} & \,\,\,\, \Sigma = 2m+M & \,\,\,\, \Delta = (M-m)/\Sigma  \nonumber \\ 
{\rm IH:} & \,\,\,\, \Sigma = m+2M & \,\,\,\, \Delta = (m-M)/\Sigma  \,.
\end{eqnarray}

 To determine the possible hierarchy is very relevant as it can complement results from neutrinoless double-$\beta$ decay to  help determine the nature of the neutrino itself \citep{Jimenez}: is it its own anti-particle? that is, is it a Majorana fermion?  
As discussed in \citet{Bahcall} (see their Fig.~3 recalling that a light neutrino mass of $0.07$eV corresponds to $\Sigma\sim 0.2-0.25$eV), if  the next generation of neutrinoless double-$\beta$ decay experiments find a signal, then neutrinos are Majorana. If these experiments do not see a signal,  it is important to discriminate if that is because the signal is below the detection threshold, or  neutrinos are truly  Dirac particles. Here is where cosmology enters \citep{Jimenez}: if $\Sigma >0.25$eV, then the hierarchy is effectively degenerate and neutrinos are Dirac. The interesting region to determine the hierarchy  is for $0.1{\rm eV}<\Sigma<0.25$eV, where the absence of  neutrinoless double-$\beta$ decay indicates that neutrinos are Dirac only if the hierarchy is inverted. 
In \citet{Jimenez} we addressed the above question using linear theory to predict the effect of neutrinos on the matter power spectrum. Our conclusion was that an ambitious survey {\it \`a la}  stage IV dark energy task force \citep{DETF}, could in principle distinguish between the inverted and normal hierarchy if $\Sigma <0.15$eV.
However, because the effect on the matter power spectrum of neutrinos extends into the non-linear regime, it is crucial to perform  N-body simulations that include massive neutrinos to properly  quantify the effect.  

The main physical effect that distinguishes different hierarchies is the fact that neutrinos of different masses have different transition  from relativistic to non-relativistic thus influencing the  shape of the matter power spectrum \citep{Slosar,debernardis}. Note that  neutrino oscillation experiments  rule out  large regions in the $\Sigma-\Delta$ parameter space (see Fig.~\ref{fig:delta_sigma}) and therefore it is worth investigating only the allowed  region (gray swath in Fig.~\ref{fig:delta_sigma}).

\section{Numerical Simulation Method}

In this Letter we study the effect of the neutrino mass splitting on the \emph{non-linear} matter power spectrum. In particular, we are interested in the question of how and if the neutrino hierarchy leaves an imprint on the matter power spectrum in the non-linear regime. To address this question we run cosmological N-body simulations for several different neutrino mass configurations: A) 3 massless neutrinos; B) 3 neutrinos of equal mass (degenerate case, $\Sigma=0.3$eV); C) 1 massless neutrino and 2 massive neutrinos (inverted hierarchy, $\Sigma=0.1$eV, $\Delta=-0.50$); D) 2 light neutrinos and 1 heavy neutrino (normal hierarchy, $\Sigma=0.1$eV, $\Delta=0.32$).


\begin{figure*}[htb]
\begin{center}
\includegraphics[trim= 20mm 20mm 20mm 20mm, clip=true ,angle=0,width=0.33\textwidth]{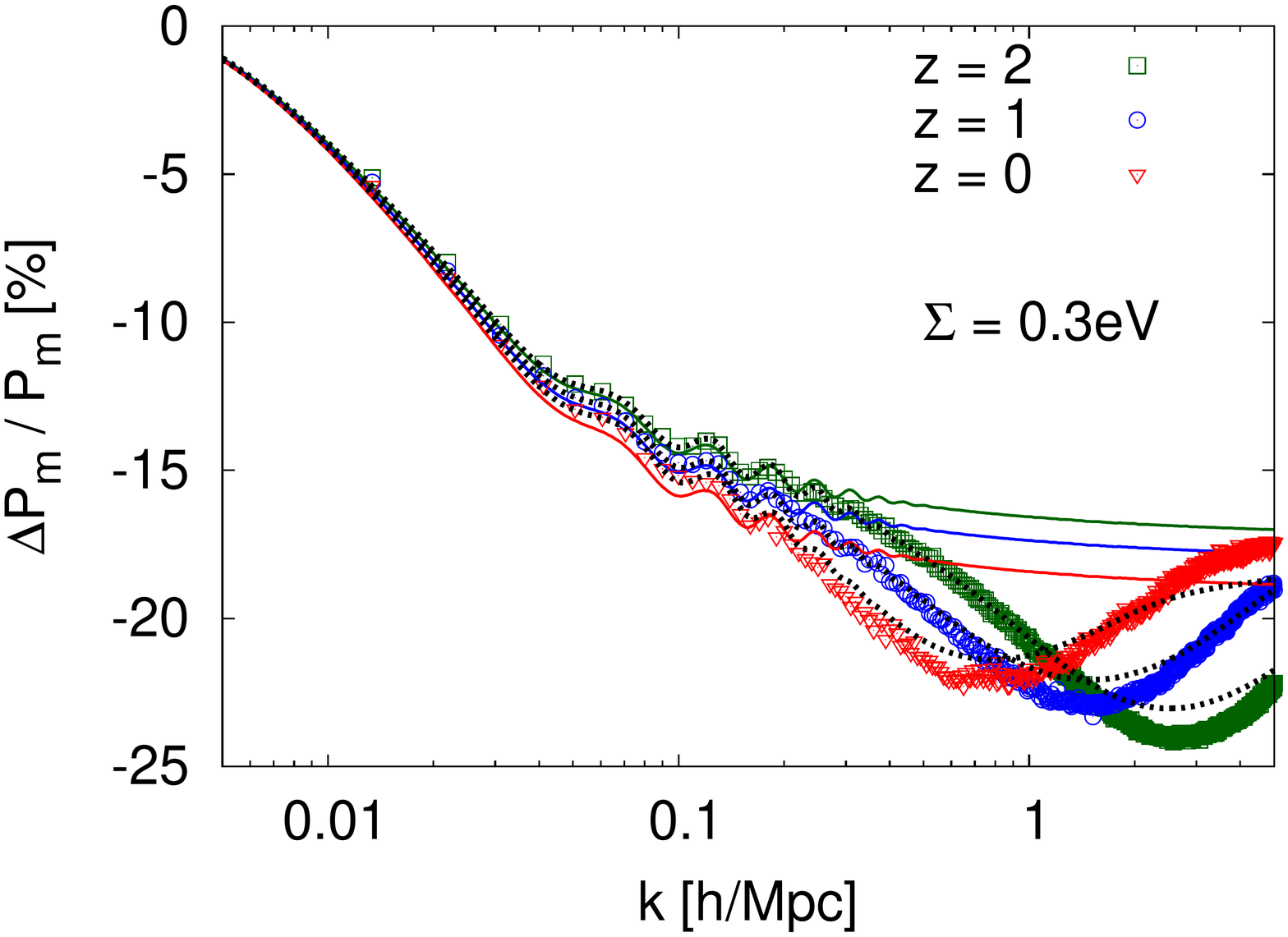}
\includegraphics[trim= 20mm 20mm 20mm 20mm, clip=true ,angle=0,width=0.33\textwidth]{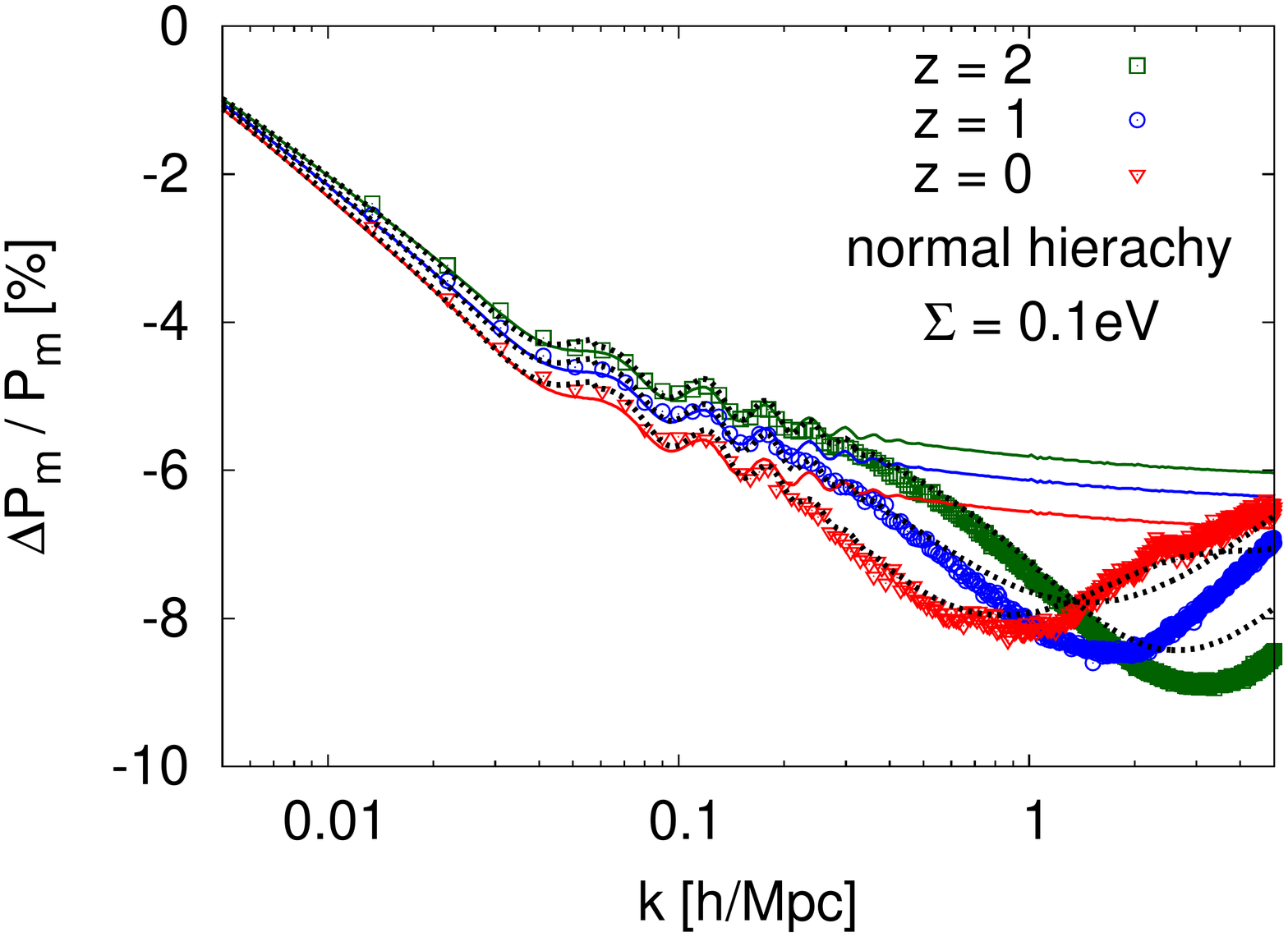}
\includegraphics[trim= 20mm 20mm 20mm 20mm, clip=true ,angle=0,width=0.33\textwidth]{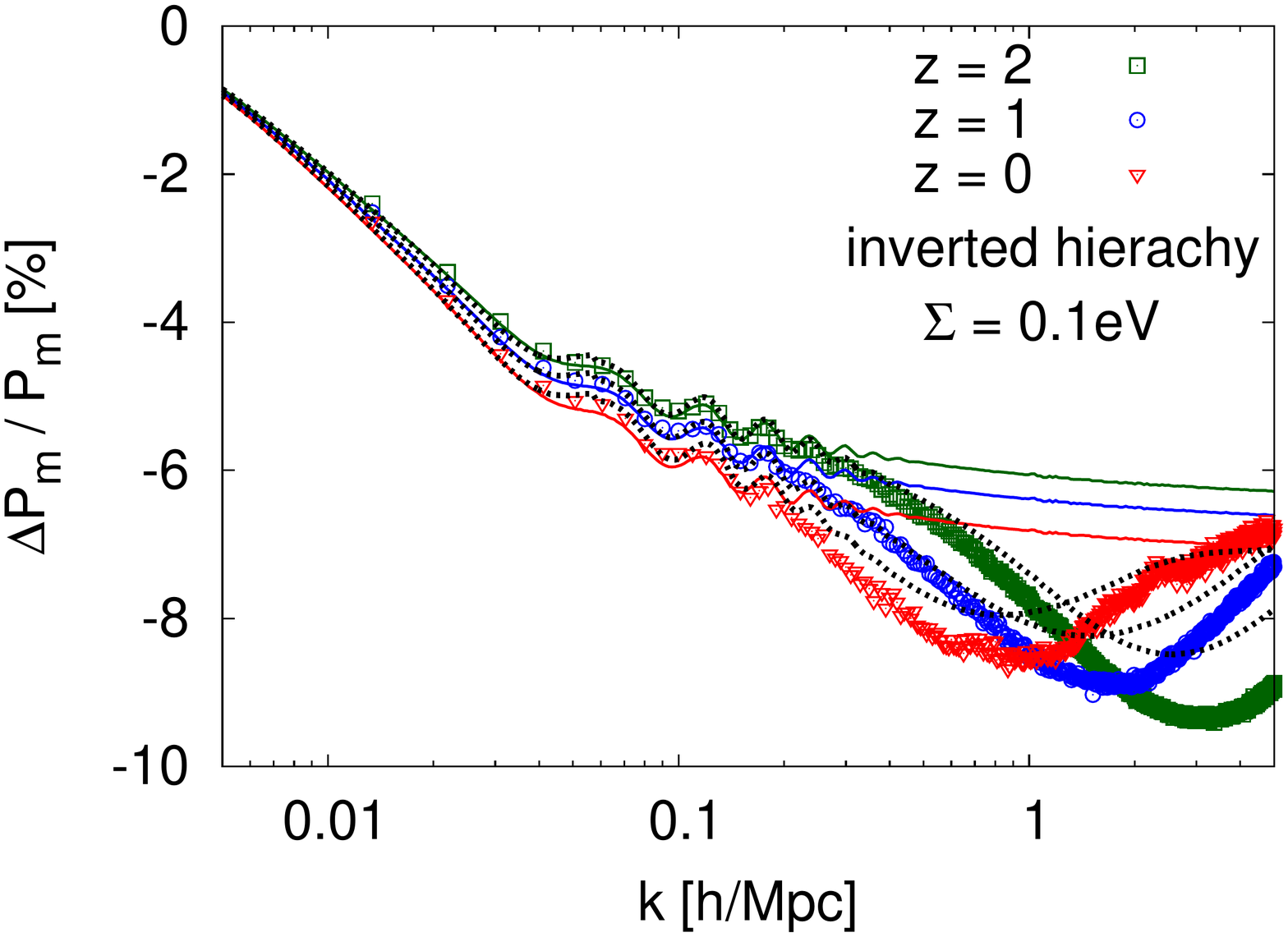}
\end{center}
\caption
{Fractional difference in the total matter power spectrum (CDM+baryons+massive neutrinos) of the massive neutrino runs and the massless neutrino run. Data points show the simulation results. The solid lines depict the linear theory predictions, while the black dotted lines show the estimated effect including non-linearities using the extended HALOFIT formula \citep{Bird}.
The left panel shows the degenerate case with $\Sigma = 0.3{\rm eV}$. The middle and right panel show the normal and inverted hierarchy, respectively, with  $\Sigma= 0.1{\rm eV}$. 
Note the typical shape of the effect in the non-linear case: non-linearities enhance the effect and lead to a maximum in the suppression located at mildly non-linear scales.
}
\label{fig:Pk_ratio_deg}
\label{fig:Pk_ratio}
\end{figure*}


So far there have been mainly two approaches to incorporating neutrinos into cosmological N-body simulations: sampling the neutrino density with particles just like in the case of cold dark matter \citep{white,klypin,Brandbyge1,Viel} or evolving the neutrino density on a grid (with a fixed size) using linear theory \citep{Brandbyge2,Viel}. These two approaches were combined into a hybrid method by \citet{Brandbyge3}.
The particle-based approach has the advantage of being able to capture the non-linear evolution of the neutrinos and their effect on the cold matter components, while the linear evolution of the neutrino density on a grid is by construction only able to model the linear gravitational effect of the neutrinos on the non-linear matter distribution. In the particle-based approach, however, the neutrinos are always treated as non-relativistic particles, since standard cosmological N-body codes do not include relativistic effects. Another problem of this approach is that the finite number of particles used to sample the neutrino density generate shot noise. As the neutrinos have high thermal velocities, they move quickly away from their initial positions and give rise to a Poisson-like shot noise.
Both these problems can be somewhat circumvented by starting the simulation at a sufficiently low redshift, when the neutrinos are practically non-relativistic and their input power spectrum is large compared to the shot noise. On the other hand, one does not want to start too late, when non-linearities have already become significant on the scales of interest. Hence, in order to keep the shot noise sub-dominant already at the initial redshift of the simulation (ideally this would be the redshift at which the neutrinos become effectively non-relativistic), a large number of neutrino tracers is required.  

Studies by \citet{Brandbyge3} and \citet{Bird} have shown that in spite of the shortcomings mentioned above the particle-based approach is more accurate in computing the effect of massive neutrinos on the non-linear matter power spectrum than the grid-based approach. This is especially the case, when looking at relative quantities like the ratio of the power spectrum with and without massive neutrinos, since in this case  systematic effects due to the rather late start of the simulation cancel out to a large extent. Hence, in this paper, we present results of particle-based simulations only.

We focus on scales in the range of $0.01\, h\,{\rm Mpc}^{-1} < k < 1\, h\,{\rm Mpc}^{-1}$. These scales will be probed to high accuracy by future surveys and are much less affected by baryonic physics (which we neglect in this paper) than smaller scales. 
Hence, we simulate a volume of $\left( 600\, h^{-1}\,{\rm Mpc} \right)^3$ with a particle load of 1 billion cold matter tracers and 2 billion neutrino tracers. 

We adopt a flat $\Lambda$CDM cosmology with cosmological parameters compatible with current observational constraints. The primordial curvature power spectrum is specified by the scalar amplitude $\Delta_\mathcal{R}^2=2.45\times 10^{-9}$ at the pivot scale $k_p=0.002\, {\rm Mpc}^{-1}$ with a spectral index $n_s=0.97$. We keep the present-day total matter fraction the same for all neutrino models: $\Omega_m=\Omega_{CDM}+\Omega_{\nu}+\Omega_b=0.27$ with the baryon fraction and the neutrino fraction given by $\Omega_b=0.046$ and $\Omega_\nu=\Sigma/(93.8{\rm eV}h^2)$, where $\Sigma$ is the sum of neutrino masses in units of eV. We choose the Hubble parameter to be $h=0.7$.
This choice of cosmological parameters yields a present-day linear mass variance in spheres of $8\, h^{-1}\,{\rm Mpc}$ of $\sigma_8\approx0.8$.

For setting up the initial conditions and for assessing the N-body simulation results, we need accurate linear predictions for the transfer functions for each component of the universe.
To this end, we use the linear Einstein-Boltzmann solver CAMB (Version Jan 2012) \citep{CAMB}, which computes the individual linear transfer functions for cold dark matter, baryons, neutrinos, and photons. As we simulate the different neutrino species separately with different neutrino tracers, we modified CAMB to store for each neutrino species a separate transfer function.

To reduce transients due to the late start of the simulation \citep{2LPT}, we implement second-order Lagrangian perturbation theory (2LPT) for the cold matter components (cold dark matter and baryons). We compute numerically the ($k$-dependent) linear growth rate from two CAMB outputs around the initial redshift. We approximate the second-order growth function $D_2$ by $D_2\approx -3/7\,D_1^2$ where $D_1$  is linear growth function \citep{Bouchet}.
At the initial redshift ($z_i=9$) the neutrino perturbations are still very much in the linear regime. Hence, for the neutrino particles, it suffices to use the Zel'dovich approximation to displace the particles from the initial grid points and to assign the gravitational velocities. Then thermal velocities drawn from the appropriate Fermi-Dirac distribution are added to the neutrino velocities. We neglect higher-order multipoles of the neutrino phase space distribution (see \citet{Ma-Bertschinger} for a treatment of the full neutrino phase space). 
Tests with CAMB where these multipoles were set to zero at $z=z_i$, and
were then evolved further to low redshift, have shown that the effect
of these initial conditions on the linear matter power spectrum is
negligible at $z \leq 2$ \citep{RdP}.

The simulations were carried out with Gadget-2 \citep{Gadget2}. We modified Gadget-2 slightly to take into account the effect of the massive neutrinos (and the radiation) on the evolution of the scale factor $a$.

Using smaller test runs, we performed convergence tests on the neutrino time stepping. We found that a maximum time step ${\rm max}(\Delta \ln a) = 0.025$ in the long-range particle-mesh force computation is sufficient for an accurate computation 
of the effect of the neutrinos on the matter power spectrum.
Note that we disabled the Courant condition for the neutrino tracers. Hence, the time step for the long-range force is determined from the velocity dispersion of the cold matter tracers alone. This speeds up the computations significantly and leaves the non-linear matter power spectrum virtually unchanged. We set the softening length of the short-range force to $20\,{\rm kpc/h}$ for both the cold matter and neutrino component.

Varying the initial redshift and the number of neutrino tracers, we confirmed that the measured ratio of the non-linear matter power spectrum with and without massive neutrinos is robust against the neutrino shot noise and the residual transients due to the late start \citep{Brandbyge1}.

We compute the total matter density contrast $\delta$ by assigning the particles to a $2048^3$ grid using the cloud-in-cell scheme. In this process cold matter and neutrino particles are weighted by the fraction they contribute to $\Omega_m$. Using Fast Fourier transforms and averaging  
$|\delta_k|^2$ over spherical shells with a bin width of $\Delta k=0.01\,h/{\rm Mpc}$, we obtain the power spectrum $P(k)=\langle |\delta_k|^2\rangle$. 
Note that we only consider the non-relativistic neutrino species sampled by particles, i.e.~the fluctuations in the radiation (relativistic neutrinos and photons) are not taken into account.
At the redshifts of interest ($z\lesssim2$), however, radiation contributes a negligible amount to the total energy budget.
One possible uncertainty in our simulation results is the shot noise coming from the cold matter particles. As the particles start off from a grid, the shot noise is sub-Poissonian and scale-dependent at high redshifts. However, for $z\le 1$ and $k<1\,h\,{\rm Mpc}^{-1}$, even a Poisson shot noise is negligible. Hence, we do not attempt to correct for shot noise.

\section{Results}
First, we compare the results of the degenerate case with $\Sigma=0.3{\rm eV}$ with previous works. We consider the fractional difference in the total matter power spectrum of models with and without massive neutrinos. The advantage of this relative quantity is that the sample variance present in the N-body simulations cancels out almost completely. 
The suppression of the total matter power spectrum due to the three degenerate massive neutrinos is shown in the left panel of Fig.~\ref{fig:Pk_ratio_deg}.
The simulation results show that non-linearities enhance the effect on mildly non-linear scales. This behavior is anticipated by perturbation theory \citep{saito,Wong,LesgourguesPT}. 
In contrast to the linear theory predictions (solid lines), in the non-linear case there is a maximum suppression, whose depth and position in $k$ depend on redshift. These numerical results are in excellent agreement with \citet{Brandbyge1} and \citet{Viel} and can be reasonably well modeled with HALOFIT \citep{Halofit}, which was extended by \citet{Bird} to model the effect of massive neutrinos (black dotted lines).

\begin{figure}[tb]
\begin{center}
\includegraphics[trim= 20mm 20mm 5mm 10mm, clip=true ,angle=0,width=0.49\textwidth]{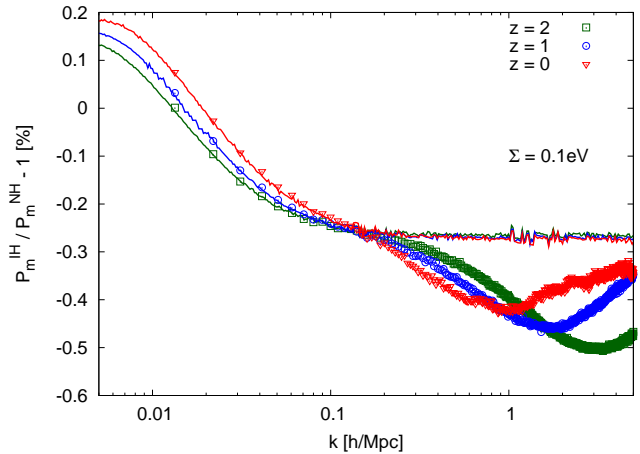}
\end{center}
\caption
{Fractional difference in the total matter power spectrum (CDM+baryons+massive neutrinos) of the inverted hierarchy and normal hierarchy run (the  sum of neutrino masses is kept fixed and only the mass splitting is varied). Note that also in this case, non-linearities enhance the effect on mildly non-linear scales compared to linear theory predictions (solid lines).
}
\label{fig:Pk_ratio_NH_IH}
\end{figure}

In the middle and right panel of Fig.~\ref{fig:Pk_ratio}, we show the power spectrum suppression for the normal and inverted hierarchy with $\Sigma=0.1$eV. We observe the same qualitative behavior as before. In order to make the small difference between the two models visible, the fractional difference in the total matter power spectrum of the two cases is shown in Fig.~\ref{fig:Pk_ratio_NH_IH}. 
Similar to the case of massive vs.~massless neutrinos, the non-linear evolution enhances the difference between the two models on mildly non-linear scales. On even smaller scales which are in the stable clustering regime, the strong non-linearities eventually overcome the initial differences between the models and the remaining effect decreases with $k$ and may eventually drop below the linear theory prediction at large $k$.

In order to test if the simulation results are affected by the realization of the initial Gaussian random field, we compare them with the results of another set of simulations with a different random realization. We find that both sets of simulation give virtually the same predictions.

\begin{figure*}[!hbt]
\begin{center}
\includegraphics[trim= 1mm 0mm 5mm 0mm, clip=true ,angle=0,width=0.33\textwidth]{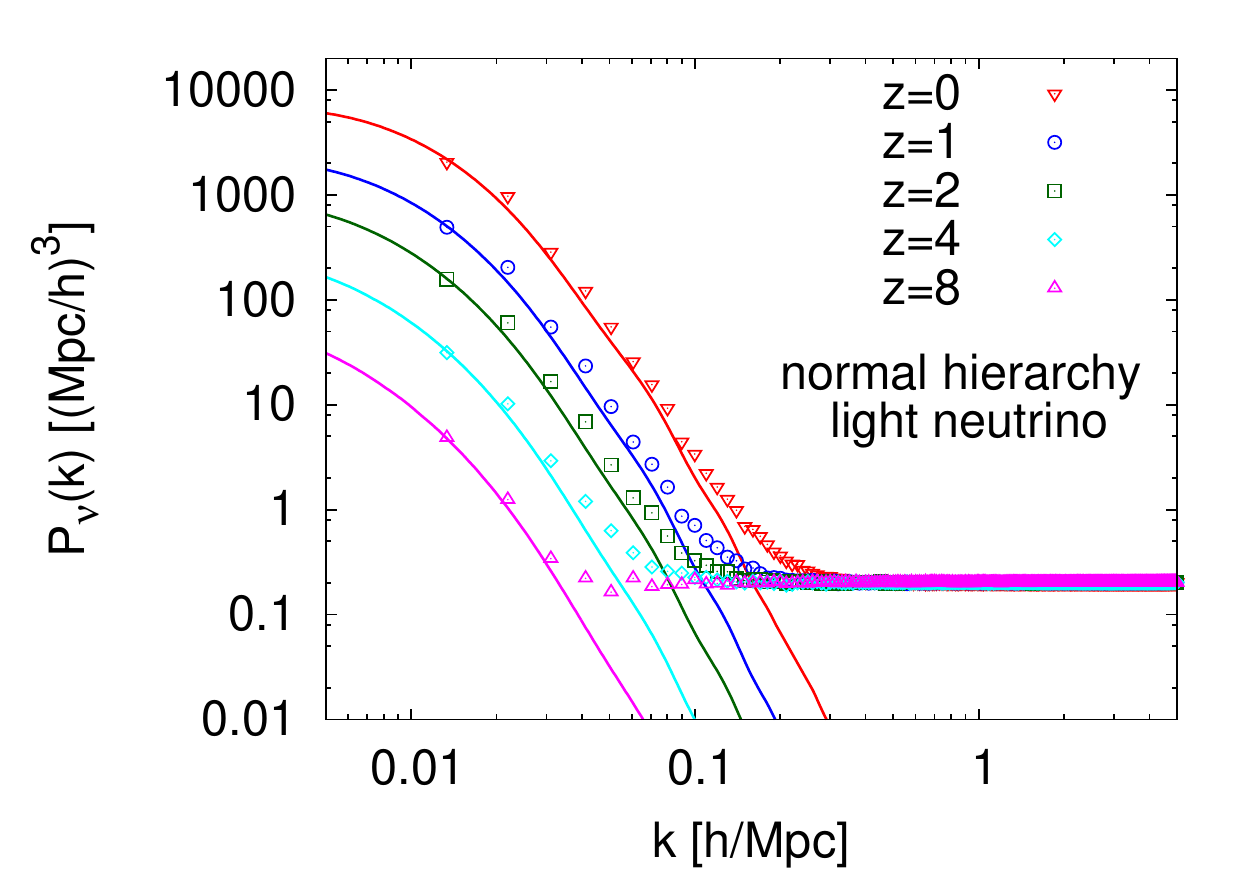}
\includegraphics[trim= 1mm 0mm 5mm 0mm, clip=true ,angle=0,width=0.33\textwidth]{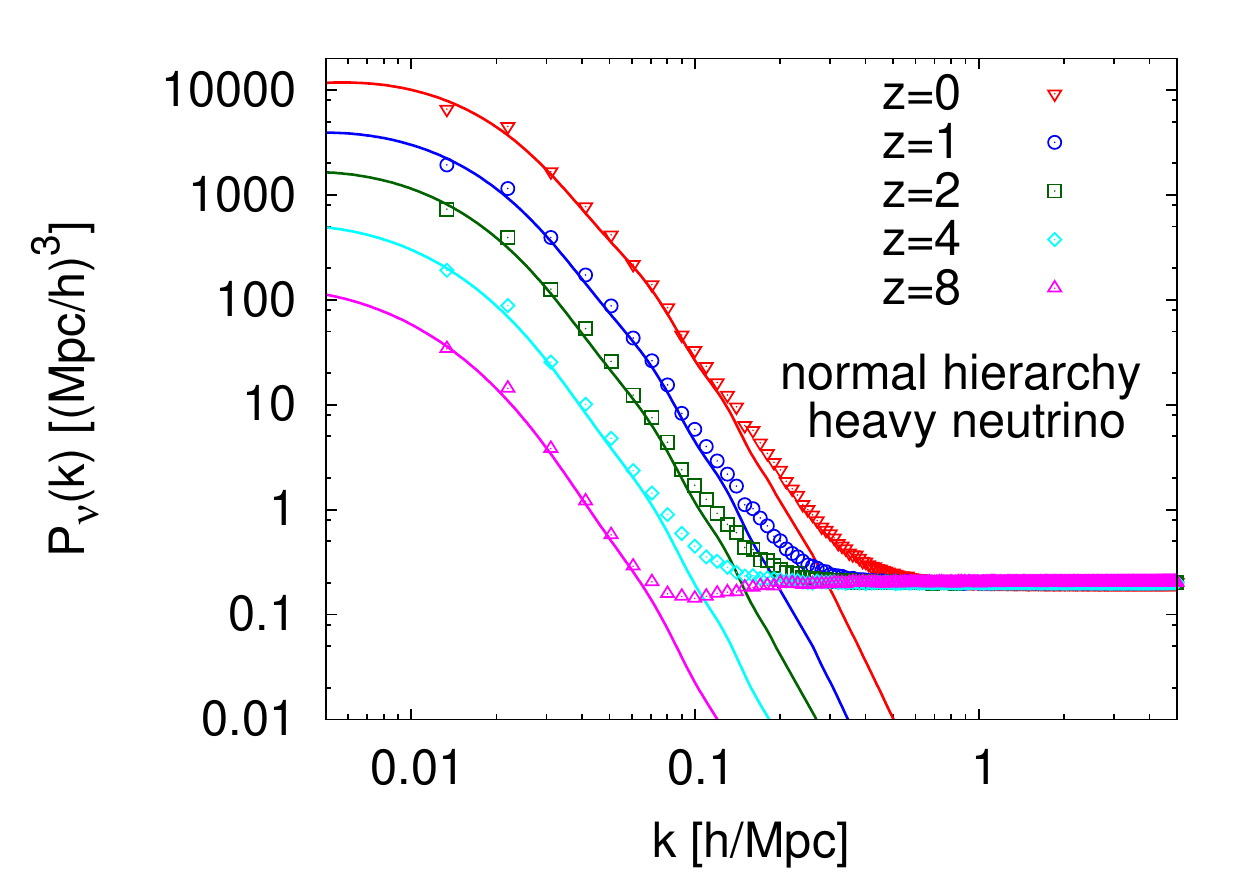}
\includegraphics[trim= 1mm 0mm 5mm 0mm, clip=true ,angle=0,width=0.33\textwidth]{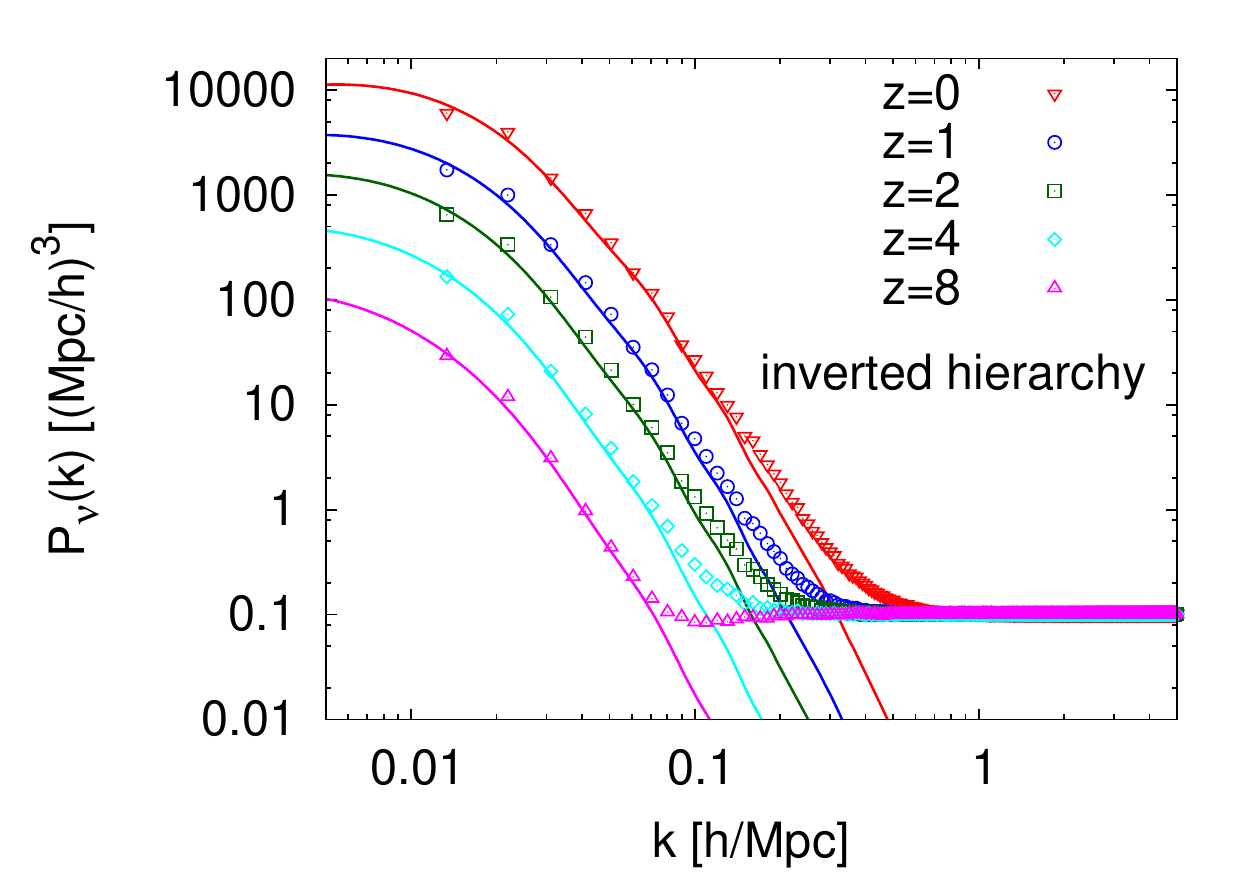}
\end{center}
\caption
{Neutrino power spectra for the normal (left panel: light neutrino, middle panel: heavy neutrino) and inverted hierarchy (right panel) at several redshifts. The shot noise effect described in the text can be easily seen. In our approximation and for the chosen value of $\Sigma$,  the normal hierarchy  has effectively two non-zero neutrino masses ($M$ and $m$ in the notation of Eq.~1), while the inverted hierarchy has effectively only one massive eigenstate $M$ (the light neutrino is massless).  
}
\label{fig:Pk_nu_IH}
\label{fig:Pk_nu_NH}
\end{figure*}

For completeness, we show the neutrino power spectra for the normal and inverted hierarchy in Fig.~\ref{fig:Pk_nu_IH}. 
On large scales ($k\lesssim 0.1\,h/{\rm Mpc}$), the neutrino power spectra from the simulations agree well with the linear predictions. 
As expected, even at high redshift we observe a Poisson-like shot noise due to the large thermal velocities of the neutrinos. 
Although this shot noise is large in the neutrino power spectrum, it is much smaller when one considers the total matter density, since neutrinos contribute less than $1\%$ to the total matter fraction.
Additionally, on scales in the non-linear regime, the matter power spectrum is dominated by the non-linear power sourced from large-scale modes, for which shot noise is negligible.

\begin{figure}[htb]
\begin{center}
\includegraphics[trim= 20mm 25mm 0mm 0mm, clip=true ,angle=0,width=0.45\textwidth]{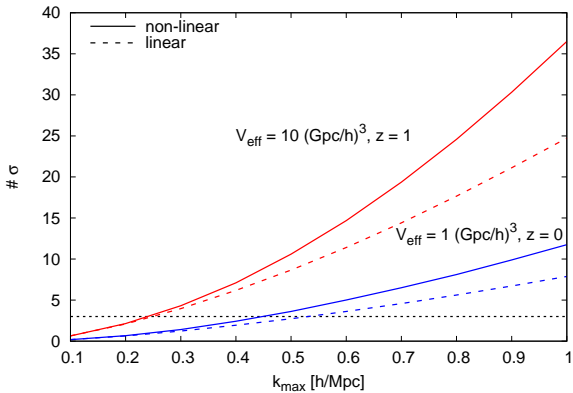}
\end{center}
\caption{Forecast of the number of sigmas separating the two hierarchies for $\Sigma=0.1$eV as a function of the maximum $k$ vector considered and for two effective volumes --$1$ and $10\, ({\rm Gpc}/h)^3$-- at $z=0$ and $z=1$ respectively. We have assumed that all other cosmological parameters are known. The horizontal dotted line indicates the $3-\sigma$  level.}
\label{fig:deltachisq}
\end{figure}

\section{Discussion and Conclusions}

The main result of our numerical experiments is that non-linearities enhance the dependence of the power spectrum on the different neutrino hierarchies, thus making the observational signature more pronounced. 
We estimate that, if all other cosmological parameters are known (including the sum of neutrino masses $\Sigma$), the two hierarchies can  be distinguished with confidence, as illustrated in Fig.~\ref{fig:deltachisq} as function of the maximum $k$ considered, making the effect potentially measurable. We have assumed an effective volume of $1\,({\rm Gpc}/h)^3$ at $z=0$ (red lines) and  $10\,({\rm Gpc}/h)^3$ at $z=1$ (blue lines).\footnote{These volumes roughly correspond  to the volume out to $z=0.5$ and  between $z=0.5$ and $z=1.5$ in $1/10$ of the sky respectively in a standard $\Lambda$CDM universe.}

Whether degeneracies with other cosmological parameters and systematic effects (galaxy bias, baryonic physics, observational limitations etc.) will cancel the detectability of the effect remains to be explored and will be considered elsewhere \citep{inprep}.
 
Our findings indicate that cosmology has the potential of determining the neutrino hierarchy in the interesting window $\Sigma \gtrsim 0.1$eV.  Signal-to-noise estimates done using  linear theory predictions  indicated that if $\Sigma$ happens to be in  the window $0.15{\rm eV}<\Sigma < 0.25$eV, cosmology could not help determine the hierarchy and thus the nature of neutrino masses \citep{Jimenez}, leaving an important gap in our knowledge of neutrino properties.  The fact that non-linearities enhance the effect compared to the linear prediction, will potentially enable cosmology to determine the hierarchy in a wider $\Sigma$ range and  possibly  close the gap. 

As an aside and already noted in \citet{Jimenez}, cosmology is  more sensitive to $|\Delta |$ than to its sign: a measurement of $|\Delta |$ in agreement with that predicted by oscillations experiments for the measured $\Sigma$ would provide a convincing consistency check for the total neutrino mass constraint from cosmology.
 
However, it is very important to keep in mind that one needs theoretical predictions of the \emph{absolute} non-linear power spectrum at least accurate to the 0.1\% level to actually be able to make any sensible measurement of the neutrino hierarchy. In this Letter we presented numerical predictions only for the relative non-linear power spectrum suppression. This relative quantity is much more robust against numerical errors. Even without massive neutrinos, it is challenging to compute the non-linear power spectrum to sub-percent precision \citep[e.g.,][]{coyote}. On small scales ($k\gtrsim 1\,h/{\rm Mpc}$), baryon physics, which is strongly model dependent and computationally very intensive, is non-negligible and makes it very hard to achieve this accuracy in the foreseeable future.
In addition, although it is in principle much less demanding to compute the linear power spectrum to high accuracy, different linear Einstein-Boltzmann codes (e.g.~CAMB \citep{CAMB} and  CLASS \citep{CLASS}) still do not agree to 0.1\% precision on the relevant scales.

Despite these very challenging and open problems, precision measurements of the large-scale structure of the universe remain an interesting avenue to determine the neutrino hierarchy.
\\

We acknowledge the participation of Carlos Pe\~na-Garay at an early stage of this project and thank him for many interesting discussions.  
CW is grateful for insightful discussions with Roland de Putter. CW and LV are supported by FP7-IDEAS-Phys.LSS 240117. LV and RJ are supported by FPA2011-29678-C02-02.


\bibliographystyle{apj}

\begin{thebibliography}{99}

\bibitem[Albrecht et al.(2006)]{DETF} Albrecht, A., 
Bernstein, G., Cahn, R., et al.\ 2006, arXiv:astro-ph/0609591 

\bibitem[Bahcall et al.(2004)]{Bahcall}
  Bahcall,  J. N., Murayama H.,  and Pena-Garay, C.,
  Phys.\ Rev.\  D { 70} (2004) 033012
  [arXiv:hep-ph/0403167].

\bibitem[Bird et al.(2012)]{Bird} Bird, S., Viel, M., 
\& Haehnelt, M.~G.\ 2012, \mnras, 420, 2551 

\bibitem[Blas et al.(2011)]{CLASS} Blas, D., Lesgourgues, J., 
\& Tram, T.\ 2011, \jcap, 7, 34 


\bibitem[Bouchet et 
al.(1995)]{Bouchet} Bouchet, F.~R., Colombi, S., Hivon, E., \& Juszkiewicz, R.\ 1995, \aap, 296, 575 


\bibitem[Brandbyge et al.(2008)]{Brandbyge1} Brandbyge, J., 
Hannestad, S., Haugb{\o}lle, T., \& Thomsen, B.\ 2008, \jcap, 8, 20 

\bibitem[Brandbyge 
\& Hannestad(2009)]{Brandbyge2} Brandbyge, J., \& Hannestad, S.\ 2009, \jcap, 5, 2 

\bibitem[Brandbyge 
\& Hannestad(2010)]{Brandbyge3} Brandbyge, J., \& Hannestad, S.\ 2010, \jcap, 1, 21 

\bibitem[Carbone et al.(2011)]{Carbone} Carbone, C., Verde, L., 
Wang, Y., \& Cimatti, A.\ 2011, \jcap, 3, 30 

\bibitem[de Bernardis et al.(2009)]{debernardis} de Bernardis, F., 
Kitching, T.~D., Heavens, A., \& Melchiorri, A.\ 2009, \prd, 80, 123509 

\bibitem[de Putter et al.(2012)]{Roland} de Putter, R., Mena, 
O., Giusarma, E., et al.\ 2012, arXiv:1201.1909

\bibitem[de Putter(2012)]{RdP} de Putter, R.\ (private communication)

\bibitem[Doroshkevich et al.(1980)]{Doro} Doroshkevich, A, G.,  Zel'dovich, Ya. B., Syunyaev, R. A.,  and  Khlopov,  M. Yu.,1980, Soviet Astron. Lett. 6, 252.

\bibitem[{Gonzalez-Garcia \& Maltoni(2008)}]{concha}
Gonzalez-Garcia, M., \& Maltoni, M. 2008, Phys.Rept., 460, 1

\bibitem[Gonzalez-Garcia et al.(2010)]{Concha2} 
Gonzalez-Garcia, M.~C., Maltoni, M., \& Salvado, J.\ 2010, arXiv:1006.3795

\bibitem[Hannestad et al.(2011)]{Hannestad} Hannestad, S., 
Haugb{\o}lle, T., \& Schultz, C.\ 2011, arXiv:1110.1257 

\bibitem[Hannestad \& Wong(2007)]{HannestadWong} Hannestad, S., \& Wong, Y.~Y.~Y.\ 2007, \emph{Journal of Cosmology and Astro-Particle Physics}, 7, 4 

\bibitem[Hannestad(2010)]{Hannestadreview} Hannestad, S.\ 2010,
  \emph{Prog. Part. Nucl. Phys.}, 65, 185

\bibitem[Heitmann et al.(2010)]{coyote} Heitmann, K., White, 
M., Wagner, C., Habib, S., \& Higdon, D.\ 2010, \apj, 715, 104 

\bibitem[Hu et al.(1998)]{HuEisensteinTegmark} Hu, W., Eisenstein, D.~J., 
\& Tegmark, M.\ 1998, \emph{Phys. Rev. Lett.}, 80, 5255 

\bibitem[{Jimenez {et~al.}(2010)Jimenez, Kitching, Pena-Garay, \&
  Verde}]{Jimenez}
Jimenez, R., Kitching, T., Pena-Garay, C., \& Verde, L. 2010, JCAP, 1005, 035

\bibitem[Kiakotou et al.(2008)]{0709.0253} Kiakotou, A., 
Elgaroy, O., \& Lahav, O.\ 2008, \emph{Phys. Rev. D}, 77, 063005 


\bibitem[Kitching et al.(2008)]{Kitching_nu} Kitching, T.~D., 
Heavens, A.~F., Verde, L., Serra, P., 
\& Melchiorri, A.\ 2008, \emph{Phys. Rev. D}, 77, 103008 

\bibitem[Klypin et al.(1993)]{klypin} Klypin, A., Holtzman, 
J., Primack, J., \& Regos, E.\ 1993, \apj, 416, 1 


\bibitem[{Komatsu {et~al.}(2011)}]{Komatsu}
Komatsu, E., {et~al.} 2011, Astrophys.J.Suppl., 192, 18

\bibitem[Lewis et al.(2000)] {CAMB} Lewis, A., Challinor, A., \& Lasenby, A.\ 2000,
  \emph{Astrophys. J.}, 538, 473


\bibitem[Lahav et al.(2010)]{LahavDES} Lahav, O., Kiakotou, A., 
Abdalla, F.~B., \& Blake, C.\ 2010, \emph{Mont. Not. Roy. Astron. Soc.}, 405, 168 

\bibitem[Lesgourgues et al.(2009)]{LesgourguesPT} Lesgourgues, J., 
Matarrese, S., Pietroni, M., \& Riotto, A.\ 2009, \jcap, 6, 17 

\bibitem[{Lesgourgues \& Pastor(2006)}]{pastor}
Lesgourgues, J., \& Pastor, S. 2006, Phys.Rept., 429, 307


\bibitem[LSST(2009)]{LSST}LSST Science Collaborations, et al.\ 2009, arXiv:0912.0201

\bibitem[Ma \& Bertschinger(1994)]{Ma-Bertschinger} Ma, C.-P., \& Bertschinger, E.\ 1994, \apj, 429, 22 



\bibitem[{{Reid} {et~al.}(2010{\natexlab{a}}){Reid}, {Verde}, {Jimenez}, \&
  {Mena}}]{Reid}
{Reid}, B.~A., {Verde}, L., {Jimenez}, R., \& {Mena}, O. 2010{\natexlab{a}},
  \jcap, 1, 3


\bibitem[Riemer--S{\o}rensen et al.(2011)]{Riemer} 
Riemer--S{\o}rensen, S., Blake, C., Parkinson, D., et al.\ 2011, 
arXiv:1112.4940

\bibitem[Saito et al.(2008)]{saito} Saito, S., Takada, M., 
\& Taruya, A.\ 2008, Physical Review Letters, 100, 191301 


\bibitem[{{Saito} {et~al.}(2011){Saito}, {Takada}, \& {Taruya}}]{Saito}
---. 2011, \prd, 83, 043529

\bibitem[Scoccimarro(1998)]{2LPT} Scoccimarro, R.\ 1998, 
\mnras, 299, 1097 


\bibitem[Slosar(2006)]{Slosar} Slosar, A.\ 2006, \prd, 73, 123501 

\bibitem[Smith et al.(2003)]{Halofit} Smith, R.~E., Peacock, 
J.~A., Jenkins, A., et al.\ 2003, \mnras, 341, 1311 

\bibitem[Springel(2005)]{Gadget2} Springel, V.\ 2005, \mnras, 
364, 1105 

\bibitem[Takada et al.(2006)]{TakadaKomatsu}Takada, M., Komatsu, E., Futamase, T. 2006,
\emph{Phys. Rev. D}, 73, 083520
 
\bibitem[Thomas et al.(2009)]{Lahavmassnu} Thomas, S.~A., Abdalla, 
F.~B., \& Lahav, O.\ 2009, \emph{Phys. Rev. Lett.}, 105, 031301

\bibitem[Viel et al.(2010)]{Viel} Viel, M., Haehnelt, M.~G., 
\& Springel, V.\ 2010, \jcap, 6, 15 


\bibitem[Wagner et al.(2012)]{inprep} Wagner, C. et al. in preparation 

\bibitem[White et al.(1983)]{white} White, S.~D.~M., Frenk, 
C.~S., \& Davis, M.\ 1983, \apjl, 274, L1 

\bibitem[Wong(2008)]{Wong} Wong, Y.~Y.~Y.\ 2008, \jcap, 10, 
35 



\end{thebibliography}

\end{document}